\documentclass[nofootinbib,aps]{revtex4}
\usepackage{amssymb,amsmath,graphicx,color,microtype,accsupp,hyperref}

\begin{document}
\title{\textit{MathGR}: a tensor and GR computation package to keep it simple}
\author{Yi Wang}
\email{tririverwangyi@gmail.com}
\affiliation{Kavli Institute for the Physics and Mathematics of the Universe (WPI),
Todai Institutes for Advanced Study, University of Tokyo,
5-1-5 Kashiwanoha, Kashiwa, Chiba 277-8583, Japan}
\affiliation{Centre for Theoretical Cosmology, DAMTP, University of Cambridge, Cambridge CB3 0WA, UK
}

\begin{abstract}
  We introduce the \textit{MathGR} package, written in \textit{Mathematica}. The package can manipulate tensor and GR calculations with either abstract or explicit indices,  simplify tensors with permutational  symmetries, convert tensors from abstract indices to partially or completely explicit indices and convert partial derivatives into total derivatives. Frequently used GR tensors and a model of FRW universe with ADM type perturbations are predefined. The package is built around the philosophy to ``keep it simple'', and makes use of latest tensor technologies of \textit{Mathematica}. 
\end{abstract}

\maketitle

\section{Introduction}

Tensor computations play a crucial role in the studies of general relativity (GR). However, in typical cases those calculations are tediously complicated. Fortunately, with the development of symbolic calculation in computer technology, a number of computer packages have been built to calculate the GR tensors, including GRTensor \cite{grtensor}, xAct \cite{xact}, Ricci \cite{ricci}, and so on. General relativitists would most likely find whatever functionality they need in those packages and work with them. 

However, in case the researchers do have a need to modify some internals of the packages, or hope to have a complete understanding of the package they use, things become complicated. Although many packages are kindly provided open source, they have in general a heavy code base. For example, the latest xAct package has over 20,000 lines of code (LOC), where the xTensor.m file alone has more than 8,000 LOC. As another example, the Ricci package has over 7,000 LOC. 

There is definitely nothing wrong with heavy weight packages which provide as many functionalities  as possible. But on the other hand, it would also be helpful to have lightweight packages which are easy for the users to understand the underlying mechanisms and modify them when needed. The \textit{MathGR} package is built for this purpose. Currently, \textit{MathGR} has about 500 LOC, and those numbers will not grow significantly in the future. The package is also built modularly, separating the tensor manipulation, GR definitions, integration by parts, specific model definition, general utilities, display parser into different files.

While keeping to be lightweight, \textit{MathGR} still provides competing functionalities for GR computations, with fast speed. The functionalities of \textit{MathGR} include
\begin{itemize}

\item Tensor simplifications with symmetries. Symmetries, anti-symmetries for any subset of indices, and any permutational symmetries defined by the \textit{Mathematica} Cycles can be brought into unique forms, such that if there can be cancellation between terms in an expression, they indeed cancel each other after simplification. 

\item Tensor calculations with either abstract or explicit (or mixed) indices, and decomposition from abstract indices to explicit ones. For example, tensors with abstract tensor indices such as the ones inside the calculation of the Ricci scalar (where the dummy indices denotes summation throughout this paper and assumed by the package)
\begin{align}
  R = - g^{\alpha\beta}g^{\gamma\delta} \partial_\gamma \partial_\delta g_{\alpha\beta} + \cdots 
\end{align}
can be calculated and simplified, and later decomposed if needed, with builtin command into
\begin{align}
  R = - g^{00}g^{00} \ddot g_{00}  - g^{00}g^{0i} \partial_i\dot g_{00} + \cdots ~,
\end{align}
where dot denotes derivative with respect to the 0 index, and $\alpha, \cdots$ run from $0$ to $n$ and $i, \cdots$ run from $1$ to $n$. Alternatively, one can also decompose the $\alpha, \cdots$ indices into completely explicit indices $0,1,\cdots, n$, or decompose $i, \cdots$ indices into explicit indices $1, \cdots, n$.

\item Simplification with total derivatives. When expanding an action of a physical system, total derivatives can either be dropped or reduced into boundary terms. \textit{MathGR} can try to reduce a given expression into total derivatives plus the rest of the terms which are minimal under various conditions \footnote{The conditions include eliminate derivatives on a certain variable, eliminate terms like $\zeta\partial_t\zeta$, or the remaining terms has minimal leaf count. However, in the leaf count case, \textit{MathGR}  cannot guarantee to give the simplest result because there is no general algorithm for such reduction as far as we know.}. For example,
\begin{align}
  a(t) \partial_t b(t) + b(t) \partial_t a(t) \quad \rightarrow \quad \partial_t [a(t) b(t)]
\end{align}

\item Cosmic perturbations. There is a builtin model calculating GR tensors for the FRW metric with ADM type perturbations. This model can on the one hand be used directly for the research of inflationary cosmology, and on the other hand provide an example for writing models.

\item Make use of \textit{Mathematica}'s builtin tensor engine. Since \textit{Mathematica} 9.0, a symbolic tensor engine is introduced. This engine is powerful and fast in reducing tensors with symmetries into unique forms. However, the interface of the tensor engine is in the coordinate independent form. The input looks like:
\begin{eqnarray} \label{internalAssumption}
\BeginAccSupp{ActualText=\detokenize{$Assumptions = (T \[Element] Arrays[{Dim, Dim, Dim}, Complexes, Antisymmetric[{1, 2}]]); TensorReduce[ TensorContract[T\[TensorProduct]T, {{1, 5}, {2, 6}, {3, 4}}] - TensorContract[T\[TensorProduct]T, {{1, 4}, {2, 6}, {3, 5}}]]}}
  && \verb!$Assumptions = ( T! 
  \in
  \verb!Arrays[{Dim, Dim, Dim}, Complexes, Antisymmetric[{1, 2}]] );! 
  \\ \nonumber
  && \verb!TensorReduce[TensorContract[T!
  \otimes
  \verb!T, {{1, 5}, {2, 6}, {3, 4}}]]! 
  \\ \nonumber && \quad
  - \verb!TensorReduce[TensorContract[T!
  \otimes
  \verb!T, {{1, 4}, {2, 6}, {3, 5}}]]!
\EndAccSupp{}
\end{eqnarray}
And the corresponding output is
\begin{eqnarray}
  \verb!Out[] = -2 TensorContract[T!
  \otimes
  \verb!T, {{1, 4}, {2, 6}, {3, 5}}]!
\end{eqnarray}

For many physicists, the above notation is not obvious, especially in case those notations come together with an expression with thousands of terms. \textit{MathGR} makes use of the above engine, but the input and output now takes the form

\begin{eqnarray}
\BeginAccSupp{ActualText=\detokenize{Simp[T[DN@"a", DN@"b", DN@"c"] T[DN@"c", DN@"a", DN@"b"] - T[DN@"a", DN@"b", DN@"c"] T[DN@"a", DN@"c", DN@"b"]] (* -2 T[DN["a"], DN["b"], DN["c"]] T[DN["a"], DN["c"], DN["b"]] *)}}
  \mathrm{Simp}[T_{abc}T_{cab} - T_{abc}T_{acb}] \qquad\rightarrow\qquad  - 2 T_{abc}T_{acb}
\EndAccSupp{}
\end{eqnarray}

On the other hand, for people using \textit{Mathematica} version lower than 9.0, \textit{MathGR} provides a simple tensor simplification engine, which can still do some calculations such as cosmic perturbations, but is not powerful enough to simplify tensor with symmetries (except totally symmetric) or permutation of indices.

\end{itemize}

With the lightweight code base and modularity, we hope \textit{MathGR} is helpful for researchers who have a demand to modify the internal functionality of a package, or those who want to take part of the internal technology from the package to build their own tools. Moreover, the development is test driven. Unit tests and integrated tests are provided thus the modifications by users are safer if all tests are passed \footnote{The unit tests and integration tests are in resources/test.m. Currently not all functions are covered by unit tests but we plan to provide better coverage in the future.}.

\section{Installation}

\textit{MathGR} can be installed with the following steps:

\begin{itemize}
\item Download the \textit{MathGR} package from
\url{https://github.com/tririver/MathGR/releases/tag/v0.1} . (Future updates will also be hosted on \url{https://github.com/tririver/MathGR} .)
\item Unpack the package.
\item Rename the directory name \verb!MathGR-0.1! (or any other version number) to \verb!MathGR!. Note that \verb!MathGR! is the directory which contains \verb!MathGR.m! together with other files.
\item Setup path so that \textit{Mathematica} can find and load \textit{MathGR}. Two options are provided here:
\begin{itemize}
\item Install: Move the \verb!MathGR! directory into a directory that \textit{Mathematica} can find by default. One can find out the default directory by evaluating \verb!$Path! in \textit{Mathematica}. Depending on operating systems, the typical directories are 
%% \begin{table}[htbp]
  \begin{center}
    \begin{tabular}{ | c | c |}
      \hline
Windows 7/Vista	& \verb!C:\Users\username\AppData\Roaming\Mathematica\Applications\! \\ \hline
Windows XP	& \verb!C:\Documents and Settings\username\Application Data\Mathematica\Applications\! \\ \hline 
Mac OS X	& \verb!~/Library/Mathematica/Applications/! \\ \hline
Linux	        & \verb!~/.Mathematica/Applications/! \\ \hline
    \end{tabular}
  \end{center}
  %% \caption{\label{tab}}
%% \end{table}

\item Use without installation: One can put \textit{MathGR} anywhere and use \verb!SetDirectory! to specify the load directory.
\end{itemize}
\end{itemize}

\section{Modules and usages}

\begin{figure}[htbp]
  \centering
  \includegraphics[width=0.9\textwidth]{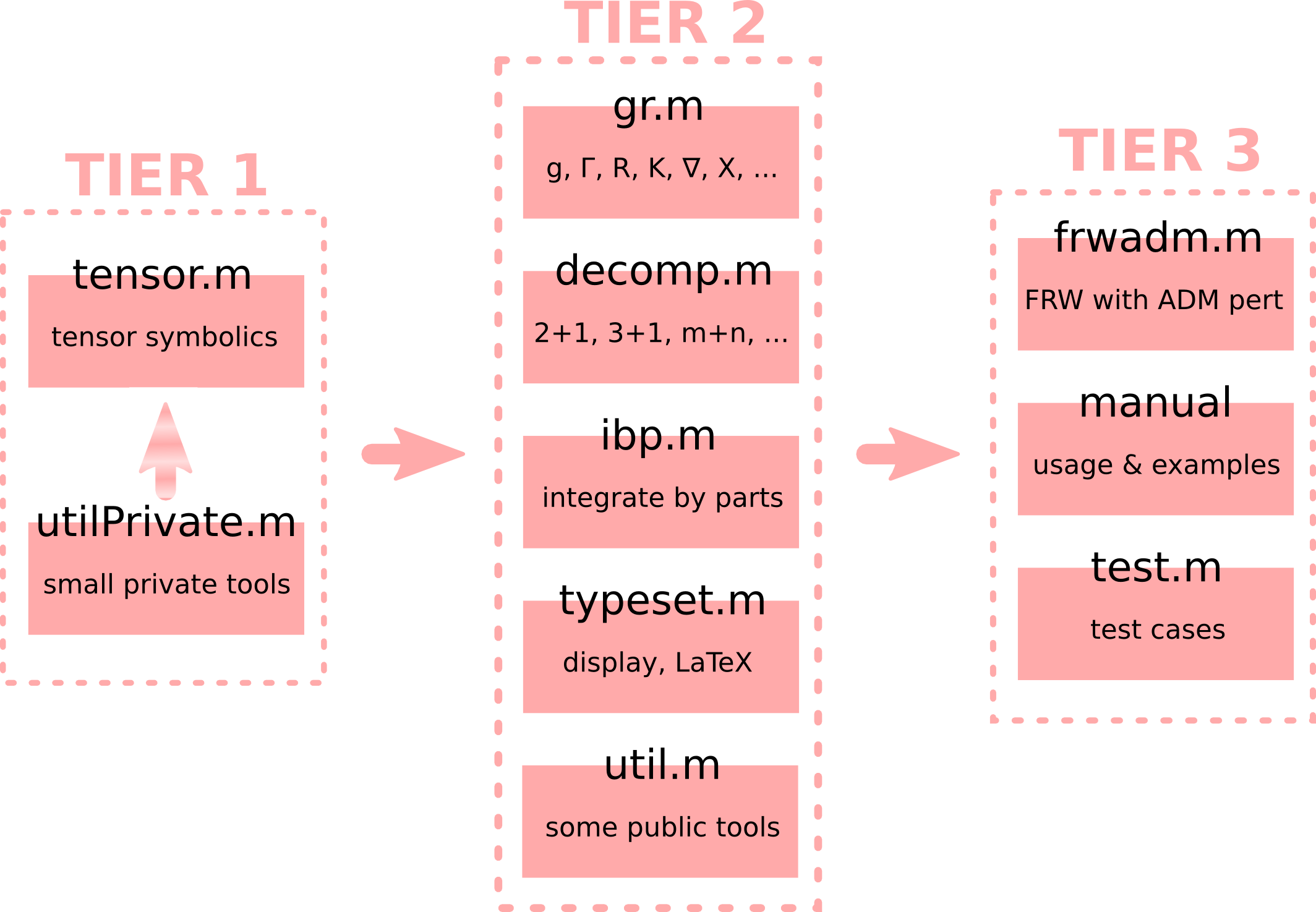}
  \caption{\label{fig:structure} The structure of the \textit{MathGR} package.}
\end{figure}

The \textit{MathGR} package is built in 3 tiers. The structure is illustrated in Fig.~\ref{fig:structure}, where Tier 2 depends on Tier 1 only, and Tier 3 can depend on anything.

In this section we introduce the functions provided by each module. To make the notation less cluttered, we have used upper and lower indices in the tensor notations. For example, the InputForm of $T^d{}_{abc}$ should be input as \footnote{There is no need to declare indices before using. Here it is not necessary to use strings (``a'', etc.) as indices. One can also use DN[$a$] where $a$ is a variable. However, once $a$ is a variable and assigned to a value like $a=b+c$, the indices will have the same replacement and will not work as desired (unless one does it by purpose). Thus it is in general recommended to use strings as indices (or Clear and Protect the indices before usage if variables are preferred).}
\begin{eqnarray} \label{tensor-input-form}
  \verb!T[UP["d"], DN["a"], DN["b"], DN["c"]]!
\end{eqnarray}
One can load the typeset.m package (to be introduced later) using 
\begin{eqnarray}
  \verb!Get["MathGR/typeset.m"]!
\end{eqnarray}
after which the InputForm \eqref{tensor-input-form} is displayed as $T^d{}_{abc}$. One may either copy this $T^d{}_{abc}$ to future input to increase readability, or keep the style of InputForm  \eqref{tensor-input-form}.

Alternatively, one may load most of MathGR packages with one command \footnote{Here and afterwards, Out[] = ... is the desired output, instead of part of input.}
\begin{eqnarray}
  &&\verb!Get["MathGR/MathGR.m"]!
  \\ \nonumber
  &&\verb!Out[] = MathGR, by Yi Wang (2013, 2014), https://github.com/tririver/MathGR. ! \\ \nonumber
  &&\verb!Bugs can be reported to https://github.com/tririver/MathGR/issues!\\ \nonumber
  &&\verb!Loaded components tensor, decomp, gr, ibp, typeset, util.!
\end{eqnarray}

\subsection{The core tensor calculations (tensor.m)}

The module tensor.m provides functions for general tensor calculation and simplification. To make use of tensor.m, we first load the package using
\begin{eqnarray}
  \verb!Get["MathGR/tensor.m"]!
\end{eqnarray}
The publicly defined functions and symbols are

\begin{itemize}
\item \textit{Define indices}\footnote{One should note that the indices can not be used in a nested way. For example, ``DN[a, UP[b]]'' is not allowed (and never necessary in GR). }: This step is optional. As we have encountered previously, we have used UP and DN as type identifiers of indices. If this is satisfying, nothing in addition needs to be done. On the other hand, one can define type identifier of indices oneself, for example
\begin{eqnarray}
  \verb!DeclareIdx[{myU, myD}, myDim, GreekIdx, Red]!
\end{eqnarray}
Here ``\{myU, myD\}'' are the new identifiers, ``myDim'' is the dimensions of the manifold on which those indices are defined. ``GreekIdx'' is a list, which shows the indices to use for dummy indices. Other choices are LatinIdx and LatinCapitalIdx\footnote{It is wise to choose different set-of-letters of indices for different type of indices. Even that we have defined class identifiers, we do not allow different class of indices to take the same index. For example f[UP[''a''], myU[``a'']] is not allowed. }. ``Red'' is the color in which the indices are typeset. Any other color that builtin in Mathematica can be chosen.  

Actually, the identifiers ``\{UP, DN\}'' are similarly defined in the package:
\begin{eqnarray}
  \verb!DeclareIdx[{UP, DN}, DefaultDim, LatinIdx, Black]!
\end{eqnarray}

As we shall cover in the next sessions, metrics can be defined associated to types of indices. Thus defining the type of indices essentially defines manifold or sub-manifold. This provides a mechanism to deal with tensor decomposition from high dimension to lower dimensions.
 
\item \textit{Declare tensor symmetry (DeclareSym)}: This step is also optional. If a tensor has no symmetry, one can use the tensor directly without declaration. On the other hand, for tensors with symmetry, one had better to declare the symmetry such that \textit{MathGR} can make use of this symmetry to simplify the tensor. 

The symmetry of a tensor $T$ can be declared using
\begin{eqnarray}
  \verb!DeclareSym[T, {list of indices}, symmetry]!
\end{eqnarray}
Here the list of indices are a sequence of UP and DN (or user defined index identifiers), representing upper and lower indices respectively. The symmetry could be either Symmetric or Antisymmetric of some slots, or some generic cyclic symmetries. The grammar for ``symmetry'' here is identical to the one defined in the \textit{Mathematica} symbolic tensor system thus one can check \textit{Mathematica} documentation for more details.

For example,
\begin{eqnarray}
  \verb!DeclareSym[T, {DN, DN, DN}, Antisymmetric[{1,2}]]!
\end{eqnarray}
defines a tensor
\begin{align}
\BeginAccSupp{ActualText={T[DN@"a", DN@"b", DN@"c"]}}
  T_{abc} ~,
\EndAccSupp{}
\end{align}

where the indices $abc$ can be any other indices, but upper and lower are distinguished here, as usual in GR. Here the tensor is anti-symmetric with permutation of $ab$ indices. In case of totally symmetric tensor (declared by Symmetric[All]), DeclareSym also sets the attribute of $T$ to be Orderless, thus the simplifications are made automatically before sending to the simplification engine.

\item \textit{Delete tensor symmetry (DeleteSym)}: The operation DeclareSym can be undone by DeleteSym. For example, 
\begin{eqnarray}
  &&\verb!DeleteSym[T, {DN, DN, DN}]!
\end{eqnarray}
removes the symmetry defined above. 

\item \textit{Simplify tensors (Simp)}: An expression (optionally with pre-defined symmetries) is brought into a unique form by the command Simp. This ensures cancellation between different terms in the expression, which can be transformed into each other by using symmetry, or with redefinition of dummy indices. For example, for the tensor $T$ defined above, one can run
  \begin{eqnarray}
    && \verb!DeclareSym[T, {DN, DN, DN}, Antisymmetric[{1,2}]]! \\ \nonumber
    && \BeginAccSupp{ActualText={T[DN@"a", DN@"b", DN@"c"]T[DN@"m", DN@"n", DN@"b"]-T[DN@"b", DN@"a", DN@"c"]T[DN@"n", DN@"m", DN@"b"]}}T_{abc}T_{mnb} - T_{bac}T_{nmb} \EndAccSupp{} \verb! // Simp!\\ \nonumber
    &&\verb!Out[] = 0!
  \end{eqnarray}
Less trivial examples are considered when introducing gr.m, with simplification of curvature tensors.

There is also a list named SimpHook. This is a list of rules that the user wants to apply before and after simplification. The rules are in the format of Rule or RuleDelayed, e.g.
  \begin{eqnarray}
    \verb!SimpHook = {! \BeginAccSupp{ActualText={T[DN@"a", DN@"b", DN@"c"]:> t F[DN@"a", DN@"b", DN@"c"]}} T_{abc}  \verb! :> !  t*F_{abc} \EndAccSupp{} \verb!}!
  \end{eqnarray}

Note that the Simp function does a lot to simplify the expressions, including making assumptions and communicating with the Mathematica command TensorReduce. If a tensor expression is expected to contain lots of terms (say, more than $10^4$) after Simp, and each term is very simple, Simp becomes then overkilling and slow. In this case, one may try to speed up by
  \begin{eqnarray}
    \verb!Simp[..., "Method"->"Fast"]!
  \end{eqnarray}
In this case dummy indices are re-arranged. However, different forms of a tensor are not guaranteed to be brought to the same form. Especially, the defined symmetries are not considered (except for the totally symmetric ones). On the other hand, the operation can be order-of-magnitude faster. 

Note that for Mathematica versions lower than 9, Simp is default to fast method because of lack of builtin function TensorReduce.

\item \textit{Partial derivative (Pd)}: Here we define the partial derivative Pd separately, instead of modifying the systemwide partial derivative D. The partial derivative can be called as
  \begin{eqnarray}
    \verb!Pd[f[indices], DN[index]]!
  \end{eqnarray}
which displayed as $\partial_\mathrm{index}f^{\cdots}{}_{\cdots}$, where $\partial$ is actually \textbackslash[CapitalSampi] in \textit{Mathematica}, which looks like $\partial$ (we try to avoid modifying system symbols such as the real $\partial$). 

The derivative Pd has linearity and Leibniz rules builtin. 

Internally, the \verb!Pd! derivative is evaluated into an object named \verb!PdT!, with
\begin{eqnarray}
  \verb!PdT[f, PdVars[a1, a2, ...]] = Pd[...Pd[Pd[f, a1], a2],...]! ~,
\end{eqnarray}
where \verb!PdVars! is a pre-defined orderless function. The reason to convert \verb!Pd! into \verb!PdT! is that, by using \verb!PdVars!, the orderlessness of partial derivative is transparent. Then it is possible to define general rules on partial derivatives.

By default, the partial derivative acting on all tensors or scalars are nonzero. To define constants with vanishing partial derivative, one has to declare explicitly. For example
\begin{eqnarray}
  \verb!PdT[f1|f2[__], _]:=0!
\end{eqnarray}
defines both f1 and f2[$\cdots$] as constants.

Note that this ``define constant'' approach is different from the builtin derivative in \textit{Mathematica}, where without explicit function dependence, the variable is by default considered as constant. We consider the former to be safer for our purpose, otherwise one may forget to define non-constants.

\item \textit{(Anti-)symmetrize tensors (Sym, AntiSym)}: The command
  \begin{eqnarray}
    \verb!Sym[expression, indices]!
  \end{eqnarray}
symmetrizes the tensor. When indices are not given, all free indices are symmetrized. For example, Sym[$f_{\mu\nu}$] gives $f_{\mu\nu}+f_{\nu\mu}$. Note that we do not add factors as $1/2$ here. The function AntiSym does similar things, only that a sign is added in front of each terms, determined by if the permutation is even or odd.

\item \textit{The kronecker $\delta$ symbol (Dta)}: The simplification of quantities like $\mathrm{Dta}_{ac}f_{bc}$ is automatic. Without the need of calling Simp, this quantity is directly evaluated into $f_{ba}$. On the other hand, if there are standalone $\mathrm{Dta}_{ab}$ which cannot be simplified, $\mathrm{Dta}_{ab}$ is replaced by $\mathrm{Hold[Dta]}_{ab}$. This HoldForm is released when doing Simp. Finally, Dta$^a{}_a$ gives DefaultDim, which denotes dimension of space or spacetime, but without predefined value.

Note that here we adopt the convention that, the Kronecker symbol's indices are not raised or lowered by the metric. For example, $\delta_{ab}$ is not $g_{ab}$, instead, $\delta_{ab}=0$ if $a\neq b$ and $\delta_{ab}=1$ if $a=b$, for any metric. This convention is used, as far as I know, widely in modified gravity and cosmic perturbations. Also, note that expressions such as $\delta_{ab}$ and $\delta^{ab}$ does not raise / lower indices (e.g. $\delta_{ab}f^a f^b \rightarrow f^a f^a$, which has the same outcome as $\delta^{ab}f^a f^b$ and $\delta^{a}{}_bf^a f^b$), 

\end{itemize}

\begin{figure}[htbp]
  \centering
  \includegraphics[width=0.6\textwidth]{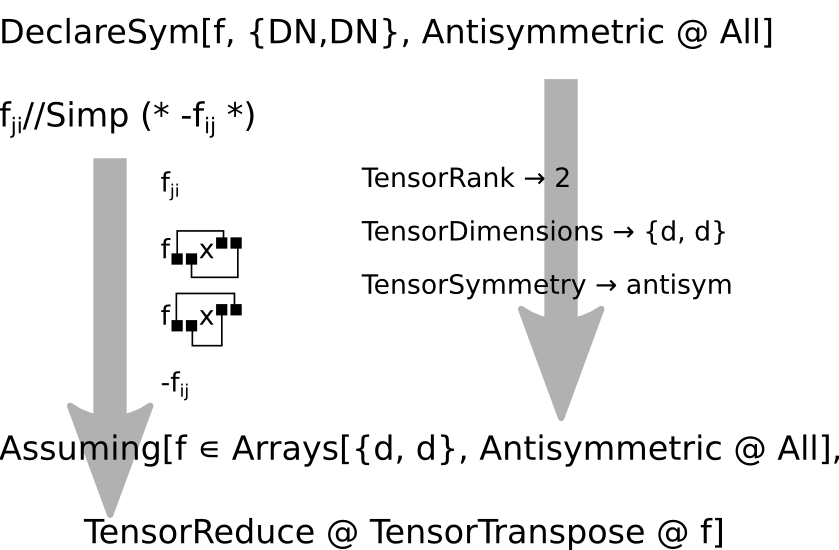}
  \caption{\label{fig:tensor} The procedure to simplify a tensor.}
\end{figure}

Before ending up this subsection, let us also briefly explain the tensor simplification mechanism under the hood. The $\verb!Simp!$ function (unless used with the $\verb!"Method"->"Fast"!$ option) calls $\verb!TensorReduce!$ to simplify tensors. And this is why at least version 9 of \textit{Mathematica} is needed when one needs the full power of tensor simplification. The procedure is described in Fig.~\ref{fig:tensor}. The tensor symmetries defined at $\verb!DeclareSym!$ are provided as assumptions of $\verb!TensorReduce!$. The tensors, on the other hand, are put into coordinate-independent abstract forms before sending into $\verb!TensorReduce!$.

\subsection{The GR definitions (gr.m)}
This package includes definitions which have made use of a metric. To load the package, use
\begin{eqnarray}
  \verb!Get["MathGR/gr.m"]!
\end{eqnarray}

\begin{itemize}
\item \textit{Metric definition and usage}: One may use a metric without any definition. However considering the symmetry and the associated inverse metric, it is convenient to define  a metric before usage. Moreover, there is also a predefined metric g, which is used by default to calculate metric contractions and curvature tensors. A metric can be defined either globally or locally. The global definition of a metric $g$ is (this metric g predefined in gr.m)
\begin{eqnarray}
    \verb!UseMetric[g, {UP, DN}]!
\end{eqnarray}
With this definition, \textit{MathGR} declares symmetry for both the metric $g_{\mu\nu}$ and the inverse metric $g^{\mu\nu}$. Also $g^\mu_\nu$ and $g^\mu_\alpha  g^\alpha_\nu$ are both aliased to $\delta^\mu_\nu$. The derivative acting on the inversed metric $\partial_\lambda g^{\mu\nu}$ is replaced using the identity $\partial_\lambda g^{\mu\nu} = - g^{\mu\alpha}g^{\nu\beta} \partial_\lambda g_{\alpha\beta}$. Also, when calculating GR quantities such as Ricci scalar R[] (to be introduced below), those quantities are calculated with respect to this metric.

When needed, one can define additional metrics, associated with other identifier of index classes (and thus on a different manifold), as
\begin{eqnarray}
  && \verb!DeclareIdx[{myU, myD}, myDim, GreekIdx, Red]!
  \nonumber \\
  && \verb!UseMetric[myG,{myU, myD}]!
\end{eqnarray}

Alternatively, one can use a metric locally, without affecting the global metric. There are two related function usages:

One can define a metric, say $h$, with all metric proprieties but without assigning the system's default metric to $h$, using
\begin{eqnarray}
    \verb!UseMetric[h, {UP, DN}, "SetAsDefault"->False]!
\end{eqnarray}

Also, one can temporarily define a metric and do some calculation with it, using, e.g.
\begin{eqnarray}
    \verb!WithMetric[h, (some tensors)]!
\end{eqnarray}
Here the (some tensors) are calculated using metric $h_{\mu\nu}$, but outside the WithMetric environment, the default metric is still $g$.

\item \textit{Curvature related symbols and tensors}: The affine connection $\Gamma^a{}_{bc}$ (named Affine in \textit{MathGR}), intrinsic curvature tensors R$^a{}_{bcd}$, R$_{abcd}$, R$^a{}_b$, R[], extrinsic curvature tensors K$_{ab}$, K[] are defined and will be calculated when called. There is also a tensor RADM[], which differs from the Ricci scalar R[] by a total derivative.

For example,
  \begin{eqnarray}
  && \verb!WithMetric[g, R[] // Simp]!
    \\ \nonumber
  &&\verb!Out[] = !\frac{3}{4} g^{ab}g^{cd}g^{ef}\partial_e g_{ac}\partial_fg_{bd}
    \verb! + 6 other terms !
  \end{eqnarray}
calculates and simplifies the Ricci Scalar with a temporary metric $g$. Note that by default, everything is calculated directly down to metrics. On the other hand, in case one wants to, for example, calculate R[] to affine connection and stops, one can locally rename the affine connection to block the calculation going on. For example,
  \begin{eqnarray}&&
    \verb!Block[{Affine=Gam}, R[] // Simp]!
    \\ \nonumber
    &&\verb!Out[] = ! -g^{ab}\mathrm{Gam}^c{}_{ad}\mathrm{Gam}^d{}_{cb} \verb! + 3 other terms!
  \end{eqnarray}

The covariant derivative is also defined, via
  \begin{eqnarray}
    \verb!CovD[expression, DN[index]]!
  \end{eqnarray}
as is standard in GR.

For example, one can check the second Bianchi identity for the curvature tensor
  \begin{eqnarray}&&
  \BeginAccSupp{ActualText={WithMetric[g, CovD[R[DN@"a", DN@"b", DN@"c", DN@"d"], DN@"e"] + CovD[R[DN@"a", DN@"b", DN@"d", DN@"e"], DN@"c"] + CovD[R[DN@"a", DN@"b", DN@"e", DN@"c"], DN@"d"]] // Simp}}
    \verb!WithMetric[g, CovD[!R_{abcd}\verb!, DN[e]] + CovD[!R_{abde}\verb!, DN[c]] + CovD[!R_{abec}\verb!, DN[d]] // Simp]!
  \EndAccSupp{} 
  \\ \nonumber
  &&\verb!Out[] = 0!
  \end{eqnarray}

\item \textit{Contraction with metric}: One can definitely use the straightforward way to do contraction. For example,
  \begin{eqnarray}&&
    \verb!f[] = g[UP[#1], UP[#2]] f[DN[#1], DN[#2]] & [Unique[], Unique[]]!
    \\ \nonumber
    &&\verb!Out[] = g[UP[$6], UP[$7]] f[DN[$6], DN[$7]]!
  \end{eqnarray}
Note that here we have introduced unique variables. (The ``\$6'' and  ``\$7'' are unique variables as returned by Mathematica, where the number could be not 6, 7 but rather any other numbers.) Otherwise there is a danger that if we use $f$ elsewhere with other tensors, the dummy indices of $f$ may coincide with the indices of other tensors and cause mistake. To make the Unique[] function easier to use, we defined Uq[n] to stand for a sequence of unique variables. Thus the above equation can be written as
  \begin{eqnarray}&&
    \verb!f[] = g[UP[#1], UP[#2]] f[DN[#1], DN[#2]] &@Uq[2]!
    \\ \nonumber
    &&\verb!Out[] = g[UP[$8], UP[$9]] f[DN[$8], DN[$9]]!
  \end{eqnarray}

However, the above expression is still long, and the length will grow quickly with lots of contractions. To save writing, we provide a function MetricContract, which is nothing fundamental, but a shorthand of writing, to allow contraction with the default metric. The above example can be rewritten as
\begin{eqnarray}&&
  \verb!f[] = f[DG[1], DG[1]] // MetricContract // Simp!
  \\ \nonumber
  &&\verb!Out[] = !f_{ab}g^{ab}
\end{eqnarray}
where DG and UG are a new type of indices, which is parsed by MetricContract. The same labels are contracted using the default metric. Here Simp is used to rewrite the dummy indices into familiar ones.

Multiple contractions can be calculated similarly, for example
\begin{eqnarray}&&
  \verb!f3[] = f[DG[1], DG[2]] f[DG[2], DG[3]] f[DG[3], DG[1]] // MetricContract // Simp!
  \\ \nonumber
  &&\verb!Out[] = ! f_{ab} f_{cd} f_{ef} g^{af} g^{bc} g^{de}
\end{eqnarray}
The above example calculates $f_a{}^bf_b{}^cf_c{}^a$, in case only $f_{ab}$ with lower indices are defined \footnote{Note that in \textit{MathGR} we do not assume relations between upper indices and lower indices (i.e. if the indices are raised and lowered by the metric, or by which metric) unless explicitly defined. This is why MetricContract is handy.}.

\end{itemize}

\subsection{Decomposition of tensors (decomp.m)}

The decomposition module can be loaded by
\begin{eqnarray}
  \verb!Get["MathGR/decomp.m"]!
\end{eqnarray}

\textit{Decomposition of tensors into lower dimensional ones (Decomp)}: This function converts the dummy indices of $m+n+\cdots$ dimensional tensors into $m$ and $n$ dimensional indices respectively. Explicit indices can also be used. The explicit indices are marked as UE[n] and DE[n] for upper and lower indices respectively, where n is a number. The usage is
\begin{eqnarray}
  \verb!Decomp[expression, rule, indices]!    
\end{eqnarray}
If indices are not given, all dummy indices are decomposed. The rules are of the form, for example,
\begin{eqnarray}
  \verb!{{DTot@#->DE@0, UTot@#->UE@0}&, {DTot@#->DN@#, UTot@#->UP@#}&}!
\end{eqnarray}
where the lower or upper index is converted into one explicit index DE[0] or UE[0], and another abstract index. There are a number of predefined decomposition schemes, namely

Decomp0i[expression, indices]: convert indices into 0 and i components. The rule is as illustrated above.

Decomp01i[expression, indices]: convert indices into 0, 1, i components.

Decomp0123[expression, indices]: convert indices into all explicit indices, 0, 1, 2, 3.

Decomp1i[expression, indices]: convert indices into 1, i components.

Decomp123[expression, indices]: convert indices into all explicit indices, 1, 2, 3.

DecompSe[expression, indices]: convert indices into two parts, where both parts has general dimensions.

Again if the indices are not explicitly specified, all dummy indices are converted. The free indices, if exist, are not touched.

In those predefined schemes, the original class identifier of indices are \{UTot, DTot\}. For example,

\begin{eqnarray}&&
  \verb!DecompSe[a[UTot["a"]] b[DTot["a"]]]!
  \\ \nonumber
  &&\verb!Out[] = a[U1["!\alpha\verb!"]]*b[D1["!\alpha\verb!"]] + a[U2["a"]]*b[D2["a"]]!
\end{eqnarray}
Here after decomposition, the indices are identified by \{U1, D1\}, and \{U2, D2\} (those index identifiers are defined in decomp.m, with dimensions Dim1 and Dim2).

As another example,
\begin{eqnarray}&&
  \verb!Decomp0i[a[UTot["a"]] b[DTot["a"]]]!
  \\ \nonumber
  &&\verb!Out[] = a[UE[0]]*b[DE[0]] + a[UP["a"]]*b[DN["a"]]!
\end{eqnarray}
Here the indices are converted into one explicit index and another abstract index.

To further ease the calculation, there is a list DecompHook, which contains a set of replacement rules to be applied after the decomposition. A frequent use case is to specify the explicit form of the higher dimensional metric using those set of rules, as illustrated in the model file frwadm.m.

One may define some ``homogeneous and isotropic background'' quantities with
\begin{eqnarray}
  \verb!PdT[b, PdVars[_DN, ___]]:=0    (* but Pd[b, DE[0]] is not zero *)!
\end{eqnarray}

\subsection{Integration by parts (ibp.m)}

One typical use case for a GR package is to expand an action
\begin{align}
  S = \int_M d^d x \sqrt{-g} ~\mathcal{L}~,
\end{align}
where if $M$ does not have boundary, total derivatives in $\mathcal{L}$ can be dropped (e.g. $\int d^3 x~dt~\partial_t f=0$). If $M$ has a boundary, those total derivatives can be reduced into boundary terms on the boundary $\partial M$.

For this purposes, we developed a module to factor $\mathcal{L}$ into total derivatives and the rest part. As usual, integration-like operations needs more intelligence than derivative-like operations. This module does not guarantee to find the total derivative for complicated cases. Nevertheless it works for simple cases and provide convenience for research. 

In case that $\mathcal{L}$ is a pure total derivative, the final target result is unique. However, typically, $\mathcal{L}$ is a total derivative plus some rest part. Here depending on use cases, we designed different criteria to try minimizing the rest part:

\begin{itemize}

\item \textit{Eliminate derivatives on a variable}: This can be used in a variation principle. For example,
  \begin{eqnarray}
    &&\verb!Get["MathGR/ibp.m"]!\\
    &&\verb!Ibp[y Pd[x, DN["i"]], "Rank" -> IbpVar[x]]!
  \end{eqnarray}
tries to eliminate derivatives acting on $x$, and gives a result
\begin{eqnarray}
  \verb!Out[] = -x Pd[y, DN["i"]] + PdHold[x y, DN["i"]]!
\end{eqnarray}
Here PdHold is a function defined to hold total derivatives. One can release this held total derivative by replacing it to Pd, or alternatively set it to zero by PdHold[\_\_]=0 if the manifold does not have a boundary.

\item \textit{Bring the rest part into standard form of a second order action}: In this case Ibp will try to eliminate terms in the rest part with more than two time derivatives on it. And the terms like $f x\dot x$ is transformed into $-\dot f x^2 /2 + $PdHold[$fx^2/2$, DE[0]]. The usage is
  \begin{eqnarray}&&
    \verb!Ibp[f x Pd[x, DE[0]], "Rank" -> IbpStd2]!
    \\ \nonumber
    &&\verb!Out[] = -(x^2*Pd[f, DE[0]])/2 + PdHold[(f*x^2)/2, DE[0]]!
  \end{eqnarray}

\item \textit{Leaf count}: If no criteria is given to Ibp, i.e. Ibp is called as Ibp[expression], the LeafCount function is applied to compare the rest part.

\end{itemize}

Internally, the Ibp function works as follows: First, a set of rules with patterns are defined for integration by parts. Then every possible rule is applied to the expression and the result is sorted. The one with simplest rest part is chosen and the same set of rules are tried repeatedly on the new result until a fixed point is reached.

It is easy to extend the above algorithm such that the rules are applied multiple times before the result is sorted and selected. However in this case the time complexity increase quickly. In typical use cases, Ibp may deal with expressions with of order 1000 or more terms (considering the complexity of the gravitational action). In this case multiple-step rules are not realistic.

Here we have introduced the Ibp function motivated by expanding an action. On the other hand, this function can be certainly used for other purposes, as long as total derivatives are wanted.

\subsection{Parsing of output (typeset.m)}

As we mentioned at the beginning of this section, the output of \textit{MathGR} can be parsed and brought into a better looking form. For this purpose, the package typeset.m should be loaded. No functions are provided in typeset.m. Instead, this module use MakeBoxes and MakeExpression to define the appearance for the tensors.

There are also predefined typeset styles for partial derivatives. Partial derivative acting on an abstract index is displayed as Capital Sampi, and time derivative (w.r.t DE[0]) is denoted by dot.

\subsection{Public utilities (util.m)}

To ease some typical calculations, some utilities are provided. Those utilities are not directly about tensor calculation but can save some writing for those calculations. The utilities can be loaded using
\begin{eqnarray}
  \verb!Get["MathGR/util.m"]!
\end{eqnarray}
and provide

\begin{itemize}

\item \textit{SolveExpr}: By default, the variable to be inputted to the \textit{Mathematica} command Solve should be atomic or a simple function. Expressions with head Plus, Times and Power are not allowed. SolveExpr solves this problem. For example, one can use
  \begin{eqnarray}&&
    \verb!SolveExpr[x^2 + y == 0, x^2]!
    \\ \nonumber
    &&\verb!Out[] = {{x^2 -> -y}}!
  \end{eqnarray}
to find a solution of $x^2$. To realize this, SolveExpr first replace $x^2$ by a unique temporary variable, solve the equation and replace the temporary variable back with $x^2$.

\item \textit{Series expansion and coefficients}: 

In \textit{MathGR}, the default variable to control orders of perturbations is named Eps. In case of a perturbation theory calculation, one multiplies every perturbation variable by Eps and expand them together (an example can be found in frwadm.m). The series expansion and extracting the coefficients simply makes use of \textit{Mathematica} functions Series and Coefficient. To save some writing, one can use
\begin{eqnarray}
  \verb!expression // SS[n]!
\end{eqnarray}
where n is an explicit integer, to expand expression up to nth order in Eps, or
\begin{eqnarray}
  \verb!expression // OO[n]!
\end{eqnarray}
to extract the order Eps$^n$ terms in the expression and disregard all other terms. Simplification function Simp is called automatically after expansion or extraction of coefficients.

%% \item \textit{Fourier transformation}: In the calculation of second order action, doing Fourier transformation can eliminate total derivatives and ease calculations such as solving gravitational constraints. For example a term is transformed as
%% \begin{align}
%%   \int d^4 x ~ \partial_i f(x) \partial_i f(x) \qquad\rightarrow\qquad \int dt \frac{d^3k}{(2\pi)^3} ~ k^2  f_\mathbf{k}(t) f_{-\mathbf{k}}(t)
%% \end{align}
%% The transformation for a second order action (and only for second order action) is provided by Fourier2[expression].

\end{itemize}

\section{A sample calculation: second order cosmic perturbations}

To illustrate an explicit use case of the package, here we calculate the cosmic perturbations of inflationary cosmology up to second order. The result is well known for decades \cite{Mukhanov:1981xt} (for a review with the same notation used here, see \cite{Wang:2013zva}). Nevertheless to present a standard and familiar calculation for illustration purpose may be more useful for a manual compared with presenting a new and unfamiliar calculation.

Here we present the input, the explanations and the final results. The intermediate outputs are long and is available in the file resources/MathGR\_Manual.nb

The model specification of a FRW universe with ADM type perturbations can be loaded by
\begin{eqnarray}
  \verb!Get["MathGR/frwadm.m"]!
\end{eqnarray}
Here the metric $g_{\mu\nu}$ is defined by
\begin{align}
  ds^2 = - N^2 dt^2 + h_{ij} (N^i dt + dx^i)(N^j dt + dx^j)~,
\end{align}
where 
\begin{align}
  N = 1+ \mathrm{Eps}~\alpha, \quad N_i = \mathrm{Eps}(b_i + \partial_i \beta) \quad h_{ij} = a^2 \exp(2 \mathrm{Eps}~ \zeta) \delta_{ij}~,
\end{align}
where $\partial_i b_i = 0$. Note that for simplicity only the scalar sector is considered in $h_{ij}$. We have yet one gauge degree of freedom in the scalar sector. We can fix the gauge by either set $\zeta=0$, or set the inflaton field to be homogeneous and isotropic. We shall consider the latter case as an example.

The action up to second order can be calculated with
\begin{eqnarray}
  && \verb!PdHold[__]:= 0 (* Total derivatives can be neglected here. *)! \\
  && \verb!Pd[!\phi\verb!|Pd[!\phi\verb!, DE[0]], _DN]:= 0 (* The inflaton perturbation is gauged away. *)! \\
 && \verb!s012 = Sqrtg (RADM[]/2 + DecompG2H[X[!\phi\verb!]] - V[!\phi\verb!]) // SS[2]!
\end{eqnarray}
Here DecompG2H is a function provided in frwadm.m (may move to more general places in case other models also need this function). This function calls Decomp0i to decompose a 4-dimensional quantity into 3+1 dimensions, where in the 4-dimensional quantities the metric $g$ is used and in the 3-dimensional quantities the metric $h$ is used. 

The background equation of motion can be derived by the first order action. For this purpose, we extract the first order action and consider the variation principle:
\begin{eqnarray}
  && \verb!s1 = s012 // OO[1]! \\
  && \verb!solBg = SolveExpr[{D[s1, !\alpha\verb!]==0, D[Ibp[s1, IbpVar[!\zeta\verb!]], !\zeta\verb!]==0}, {V[!\phi\verb!], Pd[!\phi\verb!, DE@0]^2}]! \\
  && \verb!SimpHook = Union[ SimpHook, solBg[[1]] ]!
\end{eqnarray}
where on the second line the background equation of motion is solved. Note that SolveExpr is used because we want to eliminate a composed expression $\dot\phi^2$. Also note that there are derivatives on $\zeta$ in s1. Thus we should first do integration by parts before applying the variation principle D[..., $\zeta$]=0. On the third line the background solution is added to SimpHook. Thus it will be automatically applied when simplifying the second order action.

Now we can work on the second order perturbations:
\begin{eqnarray}
  &&\verb!s2 = s012 // OO[2] // Fourier2 !\\
  &&\verb!cons = Solve[{D[s2, !\alpha\verb!]==0, D[s2, !\beta\verb!]==0, D[s2, b[DN@"a"]]==0}, {!\alpha\verb!, !\beta\verb!, b[DN@"a"]}] [[1]]! \\
  &&\verb!s2Solved = Ibp[s2 /. cons, IbpStd2]!
\end{eqnarray}
Here in s2Solved, we shall recover the well known result
\begin{eqnarray}
  \verb!Out[] = !a^3 \epsilon \dot\zeta^2  - a \epsilon  k^2\zeta^2 
\end{eqnarray}
as the second order action, with all constraints solved.

\section{Conclusion and future directions}

Here we presented the usage of a new tensor package, \textit{MathGR}, which is simple and lightweight, such that people can understand and modify the internal more easily.

We shall keep the simplicity of the package. While new functionalities are expected to be added, we shall not add functionalities which significantly increase the complexity of the package, especially for the core parts tensor.m and gr.m. 

We shall in the future add more comments to the existing code, and add broader coverage for unit tests and integrated tests. Those efforts will help for the users who want to hack and fork the package.

Finally but most importantly, as the package is being tested and used in realistic research, we expect to encounter bugs and provide bug fixes. As always, the result from \textit{MathGR} should be checked by independent calculations before being trusted and used in research.

\section*{Acknowledgments}
We thank J.~M.~Martin-Garcia for comments on an earlier version of this paper. This work was supported by the World Premier International Research Center Initiative (WPI Initiative), MEXT, Japan, a Starting Grant of the European Research Council (ERC STG grant 279617), and the Stephen Hawking Advanced Fellowship.

\end{document}